# Index wiki database:
# design and experiments[1]

*Krizhanovsky A. A.*
*St.Petersburg Institute for Informatics and Automation of RAS*

With the fantastic growth of Internet usage, information search in documents of a special type called a "wiki page" that is written using a simple markup language, has become an important problem. This paper describes the software architectural model for indexing wiki texts in three languages (Russian, English, and German) and the interaction between the software components (GATE, Lemmatizer, and Synarcher). The inverted file index database was designed using visual tool DBDesigner. The rules for parsing Wikipedia texts are illustrated by examples. Two index databases of Russian Wikipedia (RW) and Simple English Wikipedia (SEW) are built and compared. The size of RW is by order of magnitude higher than SEW (number of words, lexemes), though the growth rate of number of pages in SEW was found to be 14% higher than in Russian, and the rate of acquisition of new words in SEW lexicon was 7% higher during a period of five months (from September 2007 to February 2008). The Zipf's law was tested with both Russian and Simple Wikipedias. The entire source code of the indexing software and the generated index databases are freely available under GPL (GNU General Public License).

**INTRODUCTION**

In the USA, the 2007 nationwide survey found that more than a third of adult Internet users (36%) consulted the online encyclopedia Wikipedia [Rainie07]. The popularity of encyclopedia is probably best explained by the sheer amount of material on the site, the wide coverage of topics and the freshness of data. Wikipedia (WP) continues to gain popularity among the broad masses because it has a high rank assigned by search engines. E.g., in March 17, 2007, over 70% of the visits to Wikipedia came from search engines, according to Hitwise data [Rainie07]. More over, the search system Koru analyses Wikipedia links to expand query terms [MilneWitten07].

There are two kinds of data in Wikipedia: text and links (internal, external, interwiki, categories). Accordingly, three types of search algorithms[2] could be applied to the Wikipedia data:[3]
1. Link analysis algorithms that, in their turn, may be classified into two categories:
   - links are defined explicitly by hyperlinks (HITS [Kleinberg1999], PageRank [Brin1998], [Fortunato2005], ArcRank [Berry2003], Green [Ollivier2007], WLVM [Milne07]);
   - links are built automatically (Similarity Flooding [Melnik2002], automatic synonym extraction in a dictionary [Blondel2002], [Blondel2004], [Berry2003]);
2. Statistical text analysis (ESA [Gabrilovich2007], the similarity of short texts [Sahami06], constructing contextually similar words [Pantel2000], the self-organizing map [LeeYang02]);
3. Text and link analysis [Bharat1998], [Maguitman2005].

The earlier developed adapted HITS algorithm (AHITS) [Krizhanovsky2006a] searches for related terms by analysing Wikipedia internal links. There are many algorithms for searching related terms in Wikipedia, which can do without full text search [Krizhanovsky07a] (Table 3, p. 8). However, experimental comparison of algorithms [Gabrilovich2007], [Krizhanovsky07a] shows that the best results were obtained with the statistical text analysis algorithm ESA.

This induce us to create the publicly available index database of Wikipedia (further referred to as WikIDF[4]) and tools for database creation, which, as a whole, provides a full text search in the encyclopedia in particular, and in MediaWiki-based[5] wiki sites in general. Wikitext is text in a markup language that offers a simplified alternative to HTML.[6] Markup tags are useless for keyword search, and hence the wikitext should be converted to a text in natural language at the preliminary stage of indexing.

---
1. See Russian version of this paper: http://whinger.narod.ru/paper/ru/wiki_index_db_Krizhanovsky08.pdf. Parts of this paper are presented in the papers [Smirnov08] (chapter "Architecture"), [Smirnov08b] (in Russian, statistics of index databases, Zipf's law evaluation), [Krizhanovsky08b] (chapter "Experiments on corpus index generation").
2. We are interested in algorithms that either search for documents by keywords, or search for documents that are similar to the original one.
3. See also review of methods and applications for measuring the similarity of short texts in [Pedersen08].
4. The WikIDF abbreviation reflects the fact that the generated corpus index is suitable for TF-IDF weighting scheme.
5. MediaWiki is a web-based wiki software application used by many wiki sites, e.g. Wikipedia.

The developed software (the database and the indexing system) will allow scholars to analyse the obtained Wikipedia index database, and programmers to use this system as a part of their search engine in order to retrieve information from wiki sites.

Development of an index database and evaluation of the selected indexing method will require:
– to design the architecture of the wiki indexing system;
– to design the database (DB);
– to define regular expression rules to translate the markup into a natural language;
– to implement the software and carry out experiments.

The structure of the paper follows these steps, with a final section summarizing the results.

**ARCHITECTURE OF WIKI INDEXING SYSTEM**

In the architecture of the wiki indexing system shown in Fig. 1, interactions between the programs (GATE [Cunningham2005], Lemmatizer [Sokirko01], and Synarcher [Krizhanovsky2006a]) are presented.[7] The result produced by the system is the record level inverted index database[8], which contains a list of references to documents for each word, or rather, for each lemma. The indexing system requires three groups of input parameters:

1. The *Language* that defines the language of Wikipedia (one of 254 as of 16 Jan 2008) and the language of lemmatizing.[9] The language of WP should be defined in order to parse wikitext (see Fig. 1, function "Convert wiki-format to text" of the software module "Wikipedia Application").

2. *Database location* that is a set of parameters (host, port, login, password) for connecting to the remote database (WP and index).

3. *TF-IDF constraints* that define the size of the result index DB.[10]

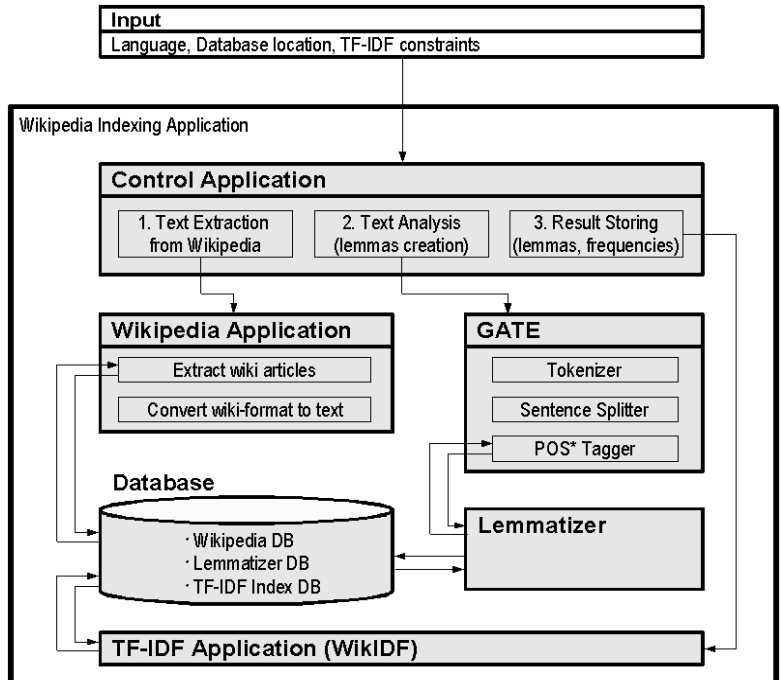

Fig. 1. Wiki indexing application architecture
(POS — Part Of Speech)

---

6  See http://en.wikipedia.org/wiki/Wikitext.
7  The fantasy about the application of this architecture for thematic wiki indexing and text filtering see in [Smirnov08].
8  See http://en.wikipedia.org/wiki/Inverted_index.
9  Since Lemmatizer has three internal databases (for Russian, English, and German).
10 E.g., the constraint of maximum number of relations *word-document* stored to the index DB. In experiments it was set equal to 1000, see Table 3.

*The Control Application* (Fig. 1) performs the following three steps for each article (wiki document).
1. The article is extracted from WP database and transformed into the text in a natural language.
2. GATE [Cunningham2005] and Lemmatizer [Sokirko01] (integrated by the interface application RussianPOSTagger)[11] generate a list of lemmas and a lemma frequency list for the text.[12]
3. The following data are stored to the index DB[13]: (i) the lemmas, (ii) the relation between the lemma and the WP article, (iii) a lemma frequency list of the article and the whole corpus.

It should be noted that two functions of the module "*Wikipedia Application*" (Fig. 1) and API for access to "*TF-IDF Index DB*" were implemented in the program Synarcher.[14] The choice of input parameters and the launching of indexing are implemented in the Synarcher module: WikIDF.[15]

**INDEX DATABASE DESIGN (TABLES AND RELATIONS)**

The relational database [Codd90] is used to store indexing information. Major factors in designing the database architecture are:[16]
– the database is created only once, the result database is read-only (so, the questions of index merge, update, maintenance, or index corruption are out of scope);
– wiki- and HTML-tags are stripped out from wikitexts; word forms are lemmatised and stored to the non-compressed database.

Number of tables in the index DB, the table's fields and relations between the tables are defined based on the problem to be solved: search for a document by the word with the help of TF-IDF formula (see below). Calculations by this formula requires three[17] tables[18] (Fig. 2)[19]:
1. *term* – the table contains lemmas of word forms (field *lemma*); number of documents, which contain word forms of the lexeme (*doc_freq*); total frequency of all word forms of this lexeme all over the corpus (*corpus_freq*);
2. *page* – a list of titles of indexed wiki articles (the field *page_title* corresponds to the field of the same name of the MediaWiki table *page*); number of words in the article (*word_count*);
3. *term_page* – the table which connects lemmas of words found in the documents with these documents.

Postfix *"_id"* in the names of tables' fields means that the field contains a unique identifier (Fig. 2). The indexed (for speed) fields are listed below the horizontal line in the frames of tables. An one-to-many relation is defined between the tables *term* and *term_page*, and between *page* and *term_page*.

---

11 See more information about Lemmatizer and RussianPOSTagger at http://rupostagger.sourceforge.net.
12 More precisely, the frequencies of all word forms (of the lexeme) found in the article (and in the corpus) are stored to the index database, where the lemma of the word form is found by the Lemmatizer program.
13 The created index DB for Russian and Simple English WP are available at: http://rupostagger.sourceforge.net, see packages *idfruwiki* and *idfsimplewiki*.
14 This functionality is available in Synarcher (from version 0.12.5), see http://synarcher.sourceforge.net.
15 WikIDF is a console application (part of Synarcher program), which depends on the RuPOSTagger program. WikIDF is bundled with Synarcher.
16 See design factors of search engine architecture: http://en.wikipedia.org/wiki/Index_(search_engine) .
17 The fourth table *related_page* is used for caching of related pages found by AHITS algorithm in previous experiments [Krizhanovsky2006a].
18 See for comparison the tables of the database in [Papadakos08] (p. 10), where the separate table is used to store the word offsets. In the paper [Papadakos08] the architecture of the Greek search engine Mitos is described.
19 The database was designed with the help of the program DBDesigner 4, see http://www.fabforce.net/dbdesigner4.

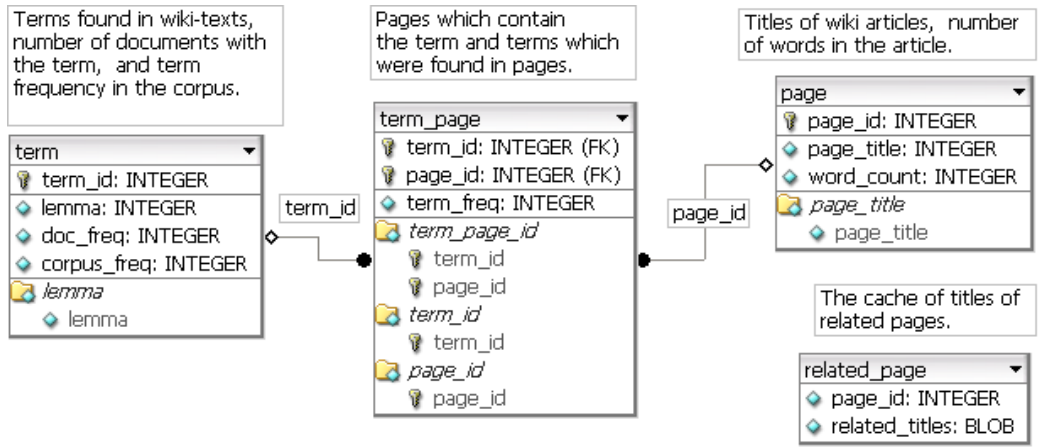

**Fig. 2. Tables and relations in the index database WikIDF**

This database allows to obtain:
- a list of lemmas of the document;[20]
- a list of documents which contain word forms of the lexeme (defined by the given lemma).[21]

Let's recall the TF-IDF formula (1), since the database was designed with the aim of fitting it. There are $D$ documents in a corpus. Given the term (lexeme) $t_i$, a document weight $w(t_i)$ can be calculated according to [Robertson2004]:

$$w(t_i) = TF_i \cdot idf(t_i) \quad ; \quad idf(t_i) = \log \frac{D}{DF_i} \quad , \tag{1}$$

where $TF_i$ is the frequency of occurrence of the term $t_i$ within a specific document (field *term_page.term_freq*, or a value of the field *term_freq* of the index database table *term_page*), $DF_i$ is the number of documents containing the term $t_i$ (field *term.doc_freq*), inverse document frequency (*idf*) serves to filter out common terms.

The *TF* count could be normalized (to prevent a bias towards longer documents) considering the number of words in the document (field *page.word_count*). Thus, the indexing database allows to calculate inverse document frequency of the term, and TF-IDF weight $w(t_i)$ of the document for a given term.

After the building of the index database it was found that the size of index is 26-38% of the size of the file with the texts to be indexed.[22, 23]

**WIKITEXT PARSING**

The articles of Wikipedia are written in wikitexts. There is a need to convert the wikitext with the aim to strip out the wiki tags and to extract the text part of them. If this step is omitted then the first hundred of the most frequent words will contain special tags like "ref", "nbsp", "br" and others.[24]

During the work, questions raised about wiki tags, whether they should be removed or parsed in another way. The questions (and parsing decisions were taken) are presented in Table 1. The most interesting (in one-line of code) cases of transformations (in Table 1) are accompanied by regular expressions [Friedl2001].

---

20 This number is constrained by user parameter *N*, so the title of *N+1* page which contains word forms of a lemma will not be stored to the table *term_page* (in experiments N is 1000).
21 There is the same constraint: the maximum number of documents is less than 1000.
22 The figures 26-38% are the ratio of columns "*Wikipedia Database*: dump, size" to "Size of archived file of index DB dump" in Table 3.
23 In the paper [Papadakos08] the result is better, namely the maximum value is 13%.
24 In fact, an analysis of the most frequent words obtained in index DB was allowed to find the tags of wikitext, which should be removed during the parsing. A code was rewritten and the database was generated again.

*The questions and decisions about parsing wikitext*

| N | Questions<br>`Source wikitext` | Answers<br>`Processed text` |
|---|---|---|
| 1 | Image caption<br>`[[Image:Asimov.jpg|thumb|180px|right|[[Isaac Asimov]] with his [[typewriter]].]]` | Save (extract text)<br>`[[Isaac Asimov]] with his [[typewriter]].` |
| 2 | Interwiki. | Save or delete (it depends on a user-defined indexing variable `b_remove_not_expand_iwiki`). |
| 3 | Category name. | Delete. |
|   | Regular expression (RE): \|\[**Category:**.+?\]\| | |
| 4 | Templates; quotes; tables. | Delete. |
| 5 | Presentational markup: italic or bold typeface.<br>`'''italic''' ''bold''` | The apostrophes are removed.<br>`italic bold` |
| 6 | Internal links<br>`[[w:Wikipedia:Interwikimedia_links|text to expand]]`<br>`[[run]]`<br>`[[Russian language|Russian]]`<br>`their [[flower|blossoms]].` | Keep the visible text, remove the hidden text.<br>`text to expand`<br>`run`<br>`Russian`<br>`their blossoms.` |
|   | RE: Internal link without pipe symbol: \|\[([^:\|]+?)\]\| | |
| 7 | External links<br>`[http://example.com Russian]`<br>`[http://www.hedpe.ru site hedpe.ru – russian fan-site]` | Keep the visible text, remove the links themselves.<br>`Russian`<br>`site – russian fan-site` |
|   | RE: Hostname (without spaces) contains the dot '.' at least once, except the last symbol:<br>(\A\|\s)\S+?[.]\S+?[^.](\|\s,!?\|\z) | |
| 8 | Footnotes.<br>`word1<ref>Ref text.</ref> – word2.` | Extract ref. text, paste it at the end of the text.<br>`word1  word2.\n\nRef text.` |

The parsing of wikitext consists of several steps that could be divided in two groups: (i) the removing and (ii) transformation of wiki tags.[25]

(i) The following tags will be removed (with the text within them):
  1. HTML comment (`<!-- ... -->`);
  2. Pre-formatted text (`<pre>...</pre>`)[26];
  3. Source code tags: `<source>` and `<code>`;

(ii) The following transformations of wiki tags are carried out:
  4. Extract the text of the reference `<ref>`, paste it at the end of the text;
  5. Remove the double curly brackets with text within them (`{{template}}`);[27]
  6. Remove tables and text (`{| table 1 \n {| A table in the table 1 \n|}|}`);
  7. Remove the accent sign in Russian texts (e.g., Па́вловск → Павловск);
  8. Remove the triple apostrophes wrapping the text (the `'''bold typeface'''` indicator), the text is kept;
  9. Remove the double apostrophes wrapping the text (the `'''cursive typeface'''` indicator), the text is kept;
  10. Extract the image caption (from the image tag), all other image tag elements are removed;
  11. Parse double square brackets (expand the internal link text, remove interwiki and category);

---

25  See the code of the function *wikipedia.text.WikiParser.convertWikiToText* in the program Synarcher.
26  Since the <pre> tags usually wrap the source code,
    see http://en.wikipedia.org/wiki/Wikipedia:How_to_edit_a_page#Character_formatting.
27  This function is called twice in order to remove {{template in {{template}}}}. The higher nest levels are not treated in the present version.

12. Parse hyperlinks enclosed in [single square brackets], link is deleted, the text is kept;
13. Remove symbols (substitute by space character) that are inadvisable input for XML parser[28]: <, >, &, "; remove also their XML-safe analogues: <, >, &, " also: ',  , –, — symbols `<br />`,`<br/>`,`<br>` will be replaced by the line break character.

This wikitext parser was implemented as one of the Java packages of the program Synarcher [Krizhanovsky2006a]. The Java regular expressions [Friedl2001] are widely used to transform elements of wikitext. The fragment of the Simple Wikipedia[29] article "Sakura" is presented in the left column of Table 2. The result of parsing this fragment taking into account all the rules (presented above) is in the right column.

**Table 2**

*An example of parsing the wikitext (Sakura)[30]*

| Source text with wiki markup tags | Parsed text |
|---|---|
| {{Taxobox<br>\| color = lightgreen<br>}}<br><br>[[Image:Castle Himeji sakura02.jpg\|thumb\|290px\|Hanami parties at [[Himeji Castle]].]] | Hanami parties at Himeji Castle. |
| '''Sakura''' or '''Cherry Blossom''' is the [[Japanese language\|Japanese]] name for decorative [[cherry]] trees, "Prunus serrulata", and their [[flower\|blossoms]]. Cherry fruit (known as "sakuranbo") come from a different species of tree. It can also be used as a name. | Sakura or Cherry Blossom is the Japanese name for decorative cherry trees, Prunus serrulata, and their blossoms. Cherry fruit (known as sakuranbo) come from a different species of tree. It can also be used as a name. |
| Sakura are object of the Japanese traditional [[custom]] of "[[Hanami]]" or "Flower viewing". | Sakura are object of the Japanese traditional custom of Hanami or Flower viewing. |
| ==See also== | ==See also== |
| * [[Hanami]] | * Hanami |
| ==Other websites== | ==Other websites== |
| * [http://shop.evanpike.com/keyword/cherry+blossom Photo Gallery of Cherry Blossoms] Sakura from Kyoto, Tokyo, Miyajima and other places around Japan | *  Photo Gallery of Cherry Blossoms Sakura from Kyoto, Tokyo, Miyajima and other places around Japan |
| [[Category:Japan]] | |

---

28 It is implied that the parser of the XML-RPC protocol of the RuPOSTagger system is used.
29 See http://simple.wikipedia.org.
30 See http://simple.wikipedia.org/wiki/Sakura.

## API INDEX WIKIPEDIA DATABASE

There are the following application programming interfaces (API) to access Wikipedia data:
- FUTEF API allows for searching within the English WP taking into account WP categories.[31] The FUTEF web-service is based on Yahoo! search engine, the search result is a Javascript object JSON;[32]
- An interface to compute the semantic relatedness of words in WP [Ponzetto07]. The Java API refers to the Perl routines (via XML-RPC) that redirect requests to the MediaWiki software.
- Two interfaces to Wikipedia and Wiktionary [Zesch08]. Experiments were carry out for extracting data from English and German Wiktionary. The main drawback is the licence "for research purposes only".
- A set of interfaces to work with Wikipedia data stored in the XML database Sedna.[33]

Since the API above (and API of Synarcher to work with MediaWiki database) are not suitable for the indexing DB, it was decided to develop a new API. Thus, an API providing access to the index Wikipedia database WikIDF has been developed.

I). The high level interface allows:[34]
  1. To get a list of terms for the given wiki article, the list is ordered by the TF-IDF weight;
  2. To get a list of articles with word forms of the lexemes (defined by the given lemmas); the articles are ordered by term frequency (TF).[35]

II). Low level functions that allow querying separate tables of the WikIDF database (Fig. 2) are implemented in the package `wikipedia.sql_idf` of the program Synarcher.

## EXPERIMENTS ON CORPUS INDEX GENERATION

The developed software for indexing wiki-texts enabled to create an index databases of Simple English Wikipedia[36] (further, denote SEW) and Russian Wikipedia[37] (RW) and to carry out experiments. The statistical data of the source / result databases and the parsing process are presented in Table 3.

In two columns ("RW / SEW 07" and "RW / SEW 08") the values of the RW parameters (at 20/09/2007 and 20/02/2008) divided by the SEW parameters (at 09/09/2007 and 14/02/2008) in 2007 and 2008 years, respectively, are presented. The parameters that characterize the Russian Wikipedia are the large quantity of lexemes (1.43 M[38]) and the total number of words in the corpus (32.93 M).

The size of Russia Wikipedia is an order of magnitude higher than Simple English one (column "RW/SEW 08"): the number of articles is 9.5 times greater, the number of lexemes is 9.6 times, the number of total words is 14.4 times.

The values in the next two columns ("SEW 08/07 %" and "RW 08/07 %") show how much the sizes of English and Russian corpuses (in comparison with itself) are increased during the five months from September 2007 to February 2008.

The last column (SEW↑ — RW↑) shows how much the rate of enlargement of the English corpus in comparison with Russian one (the difference between values of the previous two columns), namely, by 14% faster creation of new articles, and by 7% faster enriching the lexicon of Wikipedia in Simple English.

---

31 See http://api.futef.com/apidocs.html.
32 See http://json.org/.
33 See http://modis.ispras.ru/sedna/ and http://wikixmldb.dyndns.org/help/use-cases/.
34 See an example of usage of these functions in the file: synarcher/wikidf/src/wikidf/ExampleAPI.java.
35 See Table 4 with the result returned by this function (in Appendix, p. 15).
36 Most frequent 1000 words found in English Simple Wikipedia (14 Feb 2008) are listed with frequencies, see http://simple.wiktionary.org/wiki/User:AKA_MBG/English_Simple_Wikipedia_20080214_freq_wordlist.
37 Most frequent 1000 words found in Russian Wikipedia (20 Feb 2008) are listed with frequencies, see http://ru.wiktionary.org/wiki/Конкорданс:Русскоязычная_Википедия/20080220.
38 *M* symbol denotes million, k — thousand, see http://en.wikipedia.org/wiki/SI_prefix.

**Table 3**
*The statistical data of Wikipedias, parsing, and created index databases.*

| Wikipedias | Simple English (SEW 07) | Simple English (SEW 08) | Russian (RW 07) | Russian (RW 08) | RW/SEW 07 | RW/SEW 08 | SEW 08/07 % | RW 08/07 % | SEW↑ –RW↑ % |
|---|---|---|---|---|---|---|---|---|---|
| *Wikipedia Database* | | | | | | | **SEW↑** | **RW↑** | |
| Database dump, timestamp | 9 Sept 2007 | 14 Feb 2008 | 20 Sept 2007 | 20 Feb 2008 | – | – | – | – | – |
| Database dump, size, MB[39] | 15.13 | 21.11 | 240.38 | 303.59 | 15.9 | 14.4 | 140 | 126 | 13 |
| Articles, k | 19.235 | 25.22 | 204.78 | 239.29 | 10.7 | **9.5** | 131 | 117 | **14** |
| *Parsing* | | | | | | | | | |
| Total parsing, h | 3.474 | 3.628 | 52.37 | 69.26 | 15.1 | 19.1 | 104 | 132 | -28 |
| 1 page parsing, sec | 0.65 | 0.518 | 0.9206 | 1.042 | 1.42 | 2.01 | 79.7 | 113.2 | -33.5 |
| *Index Database of Wikipedia* | | | | | | | | | |
| Lexemes in the corpus, M | 0.1213 | 0.1487 | 1.2406 | **1.434** | 10.2 | **9.6** | 123 | 116 | **7** |
| Lexeme-page (<=1000 for 1 lexeme)[40], M | 1.3365 | 1.6524 | 13.49 | 15.71 | 10.1 | 9.5 | 123.6 | 116.5 | 7.2 |
| Words in the corpus, M | 1.765 | 2.2839 | 26.7 | **32.93** | 15.1 | **14.4** | 129 | 123 | 6 |
| Size of archived file of index DB dump, MB | 5.74 | 7.15 | 66.1 | 77.5 | 11.5 | 10.8 | 125 | 117 | 7 |

The following list of the computer parameters and versions of the two main programs will add a sense to the *parsing time* presented in Table 3: OS Debian 4.0 etch, Linux kernel 2.6.22.4, the AMD processor 2.6 GHz, 1 GB RAM, Java SE 1.6.0_03, MySQL 5.0.51a-3.

Now let's turn to the interesting question of corpus linguistics related to word frequency.

## ZIPF'S LAW EVALUATION ON TEXT CORPORA OF WIKIPEDIAS[41]

The empirical Zipf's law states that the frequency of the word in a corpus is inversely proportional to the rank of the word in the list of words ordered by frequency [Manning99] (p. 23). Thus the second word (in the list) will be two times less frequent than the first one, third – tree times, etc.

"One version of Zipf's law is that if we rank words in order of decreasing frequency in a large body of text, and plot a graph of the log of frequency against the log of rank, we get a straight line..." [Robertson2004]. The same graph we will plot.[42]

On Fig. 3 words are set in descending order of word frequencies (along the abscissa). Values along the ordinate indicate logarithm of the word frequencies. The curve consisting of plus symbols is based on word frequencies from Russian Wikipedia ("RW 08").[43] With the help of the method of least squares in the Scilab package [Campbell06] the following *approximation curves* (1) were calculated: $y_{100}^{RW}$ for the first hundred of the most common words in corpus (Fig. 3, dash-dot curve of the blue color) and $y_{10K}^{RW}$ for 10 thousand common words (long dash curve of the pink color).

---

39 The size of the file "...-pages-articles.xml.bz2" that contains the texts of wiki articles.
40 "Lexeme-page" is the number of relations "lexeme-page" extracted from the corpus. It could be stored (to the index DB) no more than 1000 relations for one lexeme. The number 1000 is one of the input parameters of the indexing program (See TF-IDF constraints in Fig. 1).
41 We should admit that this question attracted our attention due to the figure titled "... word frequency in Wikipedia", see http://en.wikipedia.org/wiki/Zipf%27s_law#Related_laws.
42 The source code for the calculation of an approximation and visualisation of the result in Scilab package see Appendix, p. 16.
43 "RW 08" — see explanation to this abbreviation in previous chapter and in Table 3.

$$y_{100}^{RW}(x) = \frac{e^{14.51}}{x^{0.819}} \quad ; \quad y_{10K}^{RW}(x) = \frac{e^{16.13}}{x^{1.048}} \tag{1}$$

The signs "✗" (on Fig. 3) indicate data from Simple Wikipedia "SEW 08". In the same way the following approximation curves (2) were plotted: $y_{100}^{SEW}$ (dot curve of the green color) and $y_{10K}^{SEW}$ (dash-dot-dot curve of the red color).

$$y_{100}^{SEW}(x) = \frac{e^{12.83}}{x^{0.974}} \quad ; \quad y_{10K}^{SEW}(x) = \frac{e^{14.29}}{x^{1.174}} \tag{2}$$

It should be noted that the approximation curve $y_{10K}^{RW}$ (on Fig. 3) is the more gently sloping curve (slope is −1.048) than the more steep sloping curve $y_{10K}^{SEW}$ (slope is −1.174) that corresponds to the English word frequencies decreased with faster lowering frequencies. This could be explained by several facts. Firstly, the size of Russian Wikipedia is an order of magnitude larger than Simple Wikipedia and hence a richer lexicon is used in order to explain more number of concepts. Secondly, the authors of Simple Wikipedia try to use the limited number of English words.

Fig. 3 shows that the Zipf's law holds true for texts of Wikipedias, that is the curve on the graph with a log-log scale could be approximated good enough by a straight line. At this time, the law holds better for Simple Wikipedia (0.20)[44] than for Russian Wikipedia (0.23). This could be explained by simplified language characteristics or by differences between English and Russian. A definitive answer to this question will require a solving of an industrial scale problem that is the indexing of the huge English Wikipedia.

---

44 0.20 is the difference between the slope values of approximation lines built by hundred (slope is 0.974) и ten thousand (1.174) English words.

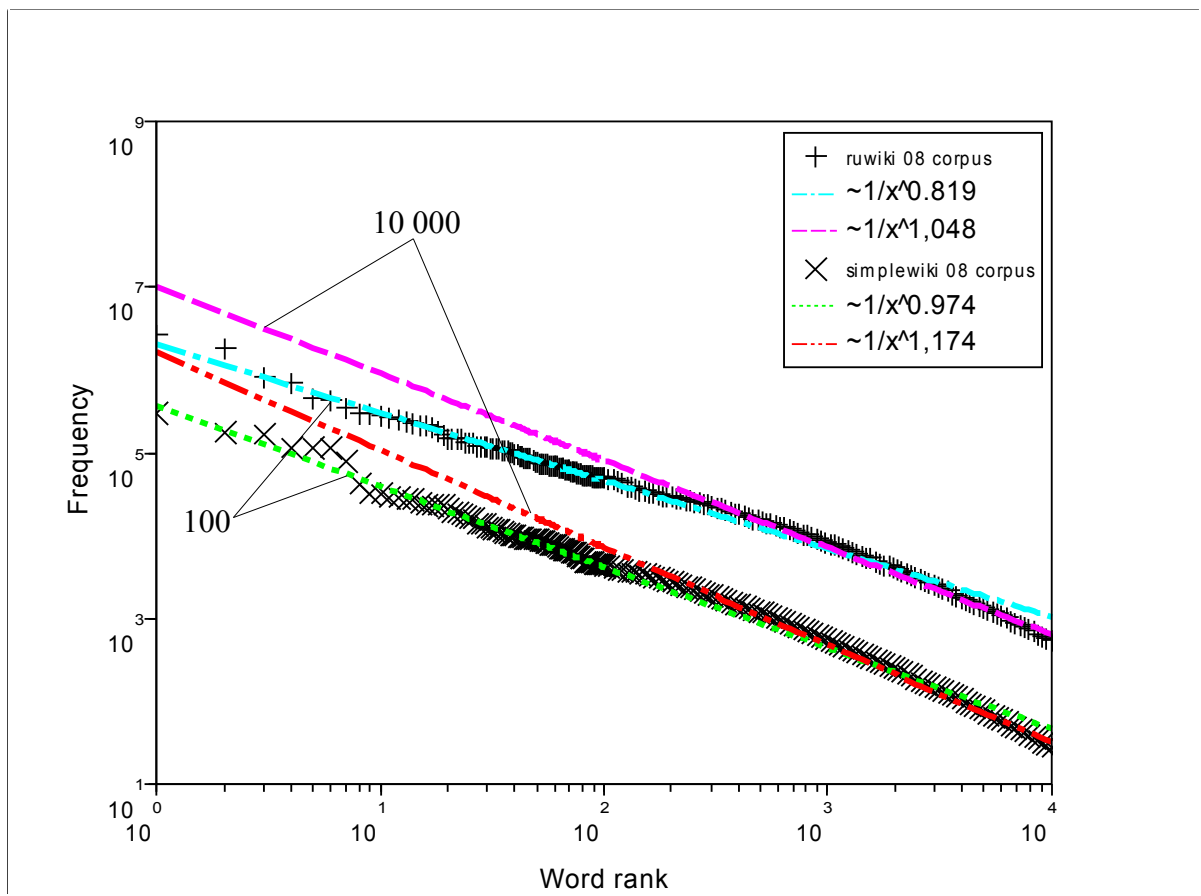

**Fig. 3. A linear dependence between the word frequency and the word rank
(the rank in the wordlist ordered by word frequency) on log-log scale
for Russian Wikipedia (ruwiki) and Simple Wikipedia (simplewiki) in February 2008,
approximation lines calculated for the 100 and 10 000 of the most common words.**

Frequency distributions of words in the two Wikipedias at two points of time are shown in Fig. 4, that is curves correspond to four corpora (index databases) presented in Table 3. Shown in Fig. 4 are the following curves (from top to down):
- *"ruwiki 08 corpus"* (dash curve of the red color) – frequency distribution of words in text corpora of Russian Wikipedia (RW) as of 20.02.2008;
- *"07 corpus"* (dot-dot-dash curve of the violet color) – word frequencies in RW as of 20.09.2007;
- *"ruwiki 07 doc"* (wide belt of the grey color) – number of documents (in RW as of 20.09.2007) which contain lexemes of the same words (on abscissa "Word rank") for which frequencies are plotted in the curve "07 corpus";
- *"simplewiki 08 corpus"* (continuous line of the violet color) – word frequencies in Simple English Wikipedia (SEW) as of 14.02.2008;
- *"simplewiki 07 corpus"* (dash curve of the green color) – word frequencies in SEW as of 09.09.2007;

The upper and lower graphics are almost the same in Fig. 4 (abscissa, logarithmic scale). The difference is that on the lower graphics, where a logarithmic scale is used on both axes. The similar behavior of the curves and an almost straight line in a log–log scale show that the Zipf's law holds for both Wikipedias during at least half a year.

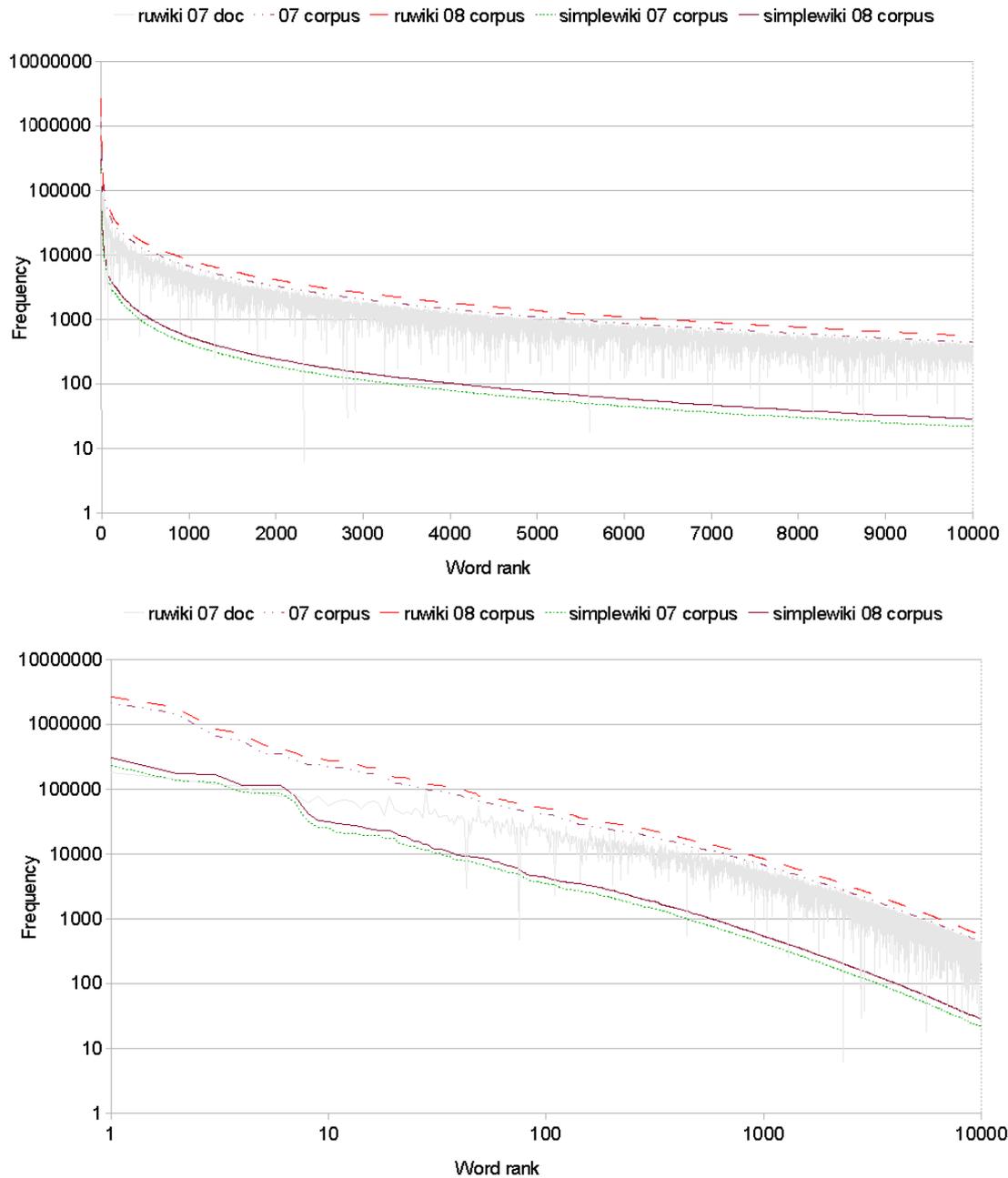

**Fig. 4 Frequencies distribution of ten thousands of the most common words in text corpora of Russian Wikipedia (ruwiki) and in Simple English Wikipedia (simplewiki) in 2007 and 2008 (upper figure) checking that the Zipf's law holds true for the corpora (lower figure).**

The following experiments based on data from the index database are left to the reader:

a) take first 1000 words sorted by frequency in a corpus, order by number of documents, draw the curve;

b) find a number of words with frequencies 1, 2, 3 ... 10 ... 1000 in a corpus (print table and draw a histogram);

c) find a number of words with length 1, 2, 3 ... 30 (table and histogram);

d) estimate the words' rank change over time (words popularity corresponds to the frequency of use), e.g. print a table with columns "word", "rank", "value of increase/decrease" (note that two index databases should be built for two Wikipedia dumps different in time); list $N$ most common words which rank has changed maximally;

e) find a number of different (unique) lexemes in a document (wiki page); estimate average and max number through all documents; find the same but normalized to the number of words in a document; list ten documents with the most rich lexicon (they contain the maximum relation between the number of unique lexemes and the number of words); solve inverse problem: list ten the most boring documents, that is long documents with the most poor lexicon.

**CONCLUSIONS**

It is not only the Internet that grows, the Wikipedia also grows up and unfolds in six-dimensional space [Geser07]:

1. Wikipedia is rapidly expanding to all major languages.
2. Total number of active collaborators is expanding.
3. The number of topics is increasing (a new group of users and a new language group may have their own interests).
4. The number of articles is rising. The articles (especially in «big» Wikipedias) have increased in the number of both the size of articles and the number of edits.
5. The internal cohesion (number of internal links, interwiki links, and categories) is increasing.
6. Wikipedia is more and more deeply "embedding" into the Web, i.e. number of external links is increasing.

Wiki is increasingly important for search engines. An indexing is an important part of a search engine work.

The architecture and implementation of the wiki texts indexing application WikIDF are presented in the paper. The interaction of the programs GATE, Lemmatizer, and Synarcher during the indexing process is described. The result of the indexing process is a list of lemmas and frequencies of lexemes stored to a database. The design of this inverted file index database is presented. The rules of converting from wiki markup to NL text are proposed and implemented in the indexing system.

The index databases of Russian Wikipedia and Simple English Wikipedia are built and compared with one another. Evaluation of the Zipf's law performance for Wikipedia texts is presented.

**ACKNOWLEDGMENTS**

This work is supported in part by grant # 08-07-00264 of the Russian Foundation for Basic Research and projects supported by the Russian Academy of Sciences (RAS) # 14.2.35 and # 1.9.

# APPENDIX 1: EXAMPLE OF RESULTS RETURNED BY THE INDEX DATABASE API

In Table 4 the data obtained using the WikIDF API are presented.[45] Table 4 contains the list of articles of Simple English Wikipedia (sorted by ΣTF) that contain both terms "green" and "tea":
- *ΣTF* – the sum of term frequencies;
- *page_title* – the title of a wiki article;
- *n_words* – the number of words in the article (page length).

**Table 4**

*The list of articles that contain both terms "green" and "tea". The list is sorted by terms summary frequencies.*

| N | ΣTF | page_title | n_words |
|---|-----|------------|---------|
| 1 | 58 | Japanese_tea_ceremony | 568 |
| 2 | 42 | Matcha | 1124 |
| 3 | 23 | Tea | 255 |
| 4 | 16 | Yerba_mate | 1628 |
| 5 | 9 | Green_tea | 59 |
| 6 | 8 | Black_tea | 81 |
| 7 | 6 | Mate_(drink) | 2143 |
| 8 | 5 | Pinhead_Gunpowder | 127 |
| 9 | 4 | Shatin_Pui_Ying_College | 2000 |
| 10 | 4 | Scientific_method | 1465 |
| 11 | 4 | List_of_plants_by_common_name | 2160 |
| 12 | 4 | Dango | 200 |
| 13 | 3 | Soba | 1169 |
| 14 | 3 | Hoboken,_New_Jersey | 1984 |
| 15 | 2 | Raspberry | 135 |
| 16 | 2 | Parfait | 121 |
| 17 | 2 | New_Zealand | 622 |
| 18 | 2 | Kumquat | 325 |
| 19 | 2 | Keane | 641 |
| 20 | 2 | Cocoa | 698 |
| 21 | 2 | 1924 | 1045 |

---

45 See "API index Wikipedia database", p. 7.

## APPENDIX 2: APPROXIMATION IN SCILAB PACKAGE

The input file consists of the word frequencies, e.g. these are first five lines of the file `ruwiki_20080220_tenK.txt`:

```
2630654
1781831
825127
698571
465278
```

The source code for (i) the calculation of a linear approximation with the help of the method of least squares and (ii) visualisation (Fig. 3) of the result in Scilab package [Campbell06] is presented below.

```
// read numbers from file to arrays
N_max=10000

// Russian Wikipedia
f=mopen('C:\experiments\ruwiki_20080220_tenK.txt','r');
for i=1:N_max
    corpus_freq=mfscanf(f,'%d');
    RU(i)=corpus_freq;
end
mclose(f);

// Simple Wikipedia
f=mopen('C:\experiments\simplewiki_20080214_only_corpusfreq10K.csv','r');
for i=1:N_max
    corpus_freq=mfscanf(f,'%d');
    SW(i)=corpus_freq;
end
mclose(f);

function z=fun(p)
    // printf(' --- start z=fun: p=%d,%d,%d\n', p(1),p(2),p(3)); // debug
    for v=1:N_max
        z(v)=log(A(v))-p(1)*log(v)-p(2);
        // printf('A(v)=%d, result z(%d)=%d\n', A(v), v, z(v)); // debug
    end
endfunction

// main two lines
//p0=[0 0];
//[ff,p]=leastsq(fun,p0);
// it's commented because:

// parameters found by leastsq by 100 and 10000 data points
p_sw100=[-0.9740246 12.826292]
p_sw10K=[-1.1740404 14.290272]

p_ru100=[-0.8186653 14.507994]
p_ru10K=[-1.0484371 16.127217]

// pars is parameter: p_sw100 or p_sw10K, etc.
function z=fn_approx(pars,v)
    z=exp(pars(1)*log(v)+pars(2));
endfunction
```

```
xbasc(); // clear the screen

// The graphic of the function with the fitted parameters
N2 = 100;
        // all points [1..N2-1] will be drawn,
        // several (5) points (thanks to log) within [N2..N_max] will be drawn
t=[1:N2-1 logspace(log10(N2),log10(N_max),100)]

plot2d(t,[SW(t)      fn_approx(p_sw100,t)'     fn_approx(p_sw10K,t)'     RU(t)
fn_approx(p_ru100,t)' fn_approx(p_ru10K,t)'],logflag="ll",axesflag=1)

legends(['ruwiki    08    corpus';'~1/x^0.819';'~1/x^1,048';'simplewiki    08
corpus';'~1/x^0.974';'~1/x^1,174'], [-1,4,6,-2,3 5],opt="ur")

a=gca();
a.x_label.text="Word rank"; // Word rank     // Номер слова
a.y_label.text="Frequency"; // Frequency     // Частота
a.font_size=2;
a.x_label.font_size=3;
a.y_label.font_size=3;

e=gce();     // legend

// a.children
// 1 - approx_ru_10K
// 2 - approx_ru100
// 3 - ruwiki
// 4 - approx_sw_10K
// 5 - approx_sw100
// 6 - simplewiki

// ruwiki
   ruwiki = a.children.children(3);
   ruwiki.thickness=1;
   ruwiki.mark_style=1;
   ruwiki.mark_mode="on";
   ruwiki.line_mode="off";
e_ru = e.children(11);
e_ru.font_size = 1;

// approx_ru100
   approx_ru100 = a.children.children(2);
   approx_ru100.thickness=3;
   approx_ru100.foreground=4;
e.children(9).font_size = 3;
e.children(10).line_style = 4;
   approx_ru100.line_style = 4;

// approx_ru_10K
   approx_ru_10K = a.children.children(1);
   approx_ru_10K.thickness=3;
e.children(7).font_size = 3;
e.children(8).line_style   = 2;    // polyline
   approx_ru_10K.line_style = 2;

// simplewiki 08 corpus

// simplewiki
   simplewiki = a.children.children(6);
   simplewiki.thickness=1;
   simplewiki.mark_style=2;
   simplewiki.mark_mode="on";
   simplewiki.line_mode="off";
e_sw = e.children(5);
e_sw.font_size = 1;
```

```
// approx_sw100
   approx_sw100 = a.children.children(5);
   approx_sw100.thickness=3;
   approx_sw100.foreground=3;
e.children(3).font_size = 3;      // text = "~1/x^0.819'"
e.children(4). line_style = 3;
   approx_sw100.line_style = 3;

// approx_sw_10K
   approx_sw_10K = a.children.children(4);
   approx_sw_10K.thickness=3;
   approx_sw_10K.foreground=5;
e.children(1).font_size = 3;      // text = "~1/x^1,048"
e.children(2).line_style = 5;     // polyline
   approx_sw_10K.line_style = 5;

drawnow();
```